\documentclass[artice,aps,pra,showpacs,twocolumn,superscriptaddress]{revtex4}
\usepackage{graphicx,color}
\usepackage{amsmath,amsthm,amsfonts,amssymb,bm}
\usepackage{times}
\usepackage{epsf}
\usepackage[colorlinks={true}]{hyperref}
\usepackage[T1]{fontenc}
\usepackage[utf8]{inputenc}
\hypersetup{citecolor={blue}, filecolor={blue}, linkcolor={blue}, urlcolor={blue}}

\newcommand{\commentold}[1]{}
\DeclareMathSymbol{:}{\mathpunct}{operators}{"3A}
\usepackage[utf8]{inputenc}
\usepackage[english]{babel}
\usepackage{amsthm}
\theoremstyle{definition}

\begin{document}

\title{Quantum memory assisted entropic uncertainty relation with moving quantum memory inside a leaky cavity}
\author{S. Haseli}
\email{soroush.haseli@uut.ac.ir}
\affiliation{Faculty of Physics, Urmia University of Technology, Urmia, Iran
}


\date{\today}

\begin{abstract}
Uncertainty principle has a fundamental role in quantum theory. This principle states that it is not possible to measure two incompatible observers simultaneously. The uncertainty principle  is expressed logically in terms of standard deviation of the measured observables. In quantum information  it has been shown that the uncertainty principle can be expressed by means of Shannon's entropy. Entropic uncertainty relation can be improve by consider an additional particle as the quantum memory. In the presence of quantum memory the entropic uncertainty relation is called quantum memory assisted entropic uncertainty relation. In this work we will consider the case in which the quantum memory moves inside the leaky cavity. We will show that by increasing  the velocity of the quantum memory the entropic uncertainty lower bound decreased. 
\end{abstract}

\maketitle
\section{Introduction}\label{Sec1}
In the classical world the measurement error is due to the inaccuracy of the measuring device. While in quantum theory, it is not possible to measure two incompatible observables simultaneously, even if the measurement instrument is accurate. It is because of the uncertainty principle in quantum theory. The first uncertainty relation was proposed by Heisenberg \cite{Heisenberg}. The Heisenberg uncertainty relation is related to momentum and position observables which are two incompatible observables. Kennard \cite{Kennard} formalized  Heisenberg uncertainty principle as $\Delta \hat{x}\Delta \hat{p}_x \geq \hbar/2$, where $\Delta \hat{x}$ and $\Delta \hat{p}_x $ are standard deviations of the position $\hat{x}$
momentum $\hat{p}_x$, respectively. Robertson \cite{Robertson} and  Schr$\ddot{o}$dinger \cite{Schrodinger} generalized Heisenberg  uncertainty relation for arbitrary two incompatible observables $\hat{Q}$ and $\hat{R}$. Based on their results, one has the following relation for arbitrary quantum state $\vert \psi \rangle$
\begin{equation}\label{hayzen}
\Delta \hat{Q} \Delta \hat{R} \geq \frac{1}{2} \vert \langle \psi \vert [ \hat{Q}, \hat{R} ] \vert \psi\rangle \vert,
\end{equation}
where $\Delta \hat{Q} = \sqrt{ \langle \psi \vert \hat{Q}^{2} \vert \psi \rangle - \langle \psi \vert \hat{Q} \vert \psi \rangle^{2}}$ and $\Delta \hat{R} = \sqrt{ \langle \psi \vert \hat{R}^{2} \vert \psi \rangle - \langle \psi \vert \hat{R} \vert \psi \rangle^{2}}$ are the standard deviations of the observables $\hat{Q}$ and $\hat{R}$, respectively, and $\left[ \hat{Q},\hat{R} \right] = \hat{Q}~\hat{R}-\hat{R}~\hat{Q}$. The left hand side (l.h.s) of Eq.(\ref{hayzen}) is called uncertainty
and the right hand side (r.h.s) is called uncertainty lower bound. This uncertainty lower bound is
depend on  quantum state $\vert \psi \rangle$, which
can be lead to a trivial bound when the commutator has zero expectation value. To avoid this drawback and to consider the notion of uncertainty as the (lack of) knowledge about a measurement outcome, it has  been suggested to use the Shannon entropy, in an information theoretical
framework, as the measure of uncertainty. The important version of entropic uncertainty relation (EUR) was proposed by Kraus \cite{Kraus} and then proved by Maassen and Uffink \cite{Maassen}. According to their results, one has the following EUR for
the state $\rho_A$ on a finite dimension Hilbert space $\mathcal{H}_A$
\begin{equation}\label{entropic1}
H(\hat{Q})+H(\hat{R}) \geq \log_{2}\frac{1}{c},
\end{equation} 
where $\hat{Q}=\lbrace\vert q_i \rangle \rbrace$ and $\hat{R}=\lbrace \vert r_j \rangle \rbrace$ represents the orthonormal bases on $\mathcal{H}_A$, $H(\hat{O})=-\sum_{o} p_{o} \log_{2} p_{o}$ is the Shannon entropy of the measured observable $\hat{O} \in \lbrace \hat{Q}, \hat{R}\rbrace$, $p_o$  is the probability distribution of the outcome $o$, and $c=\max_{\lbrace i,j\rbrace} \vert \langle q_i \vert r_j \rangle \vert^{2}$ defines the complementarity between the observables. The above EUR  provides a constant lower bound compared with Heisenberg' s uncertainty relation, because complementarity between the observables does not depend on particular states. The entropic uncertainty lower bound in Eq.(\ref{entropic1}) was improved for mixed state $\rho_A$ as 
\begin{equation}\label{entropic2}
H(Q)+H(R) \geq \log_2 \frac{1}{c}+S(\rho_A)
\end{equation} 
where $S(\rho_A)=-tr(\rho_A \log_2 \rho_A)$ is the von Neumann entropy of the state $\rho_A$. Berta et al. \cite{Berta}, generalized the EUR in Eq.(\ref{entropic2}) by considering additional particle $B$ as a quantum memory. It is  supposed that Bob have access to a particle $B$ and Alice have access to a particle $A$. Berta et al. showed that, Bob's uncertainty about the outcome of measurements $\hat{Q}$ and $\hat{R}$, is bounded by
\begin{equation}\label{berta}
S(\hat{Q} \vert B)+ S(\hat{R} \vert B) \geq \log_2 \frac{1}{c} + S(A \vert B),
\end{equation}
where $S(\hat{P} \vert B)=S(\rho^{PB})-S(\rho^{B})$ is  the conditional von Neumann entropy of the post measurement state 
\begin{equation}
\rho^{PB}=\sum_{i} (\vert p_i \rangle \langle p_i \vert \otimes \mathbb{I}) \rho^{AB} (\vert p_i \rangle \langle p_i \vert\otimes \mathbb{I}),
\end{equation}
which $ \vert p_i \rangle$ is the eigenstate of the observable $\hat{P}$, and $\mathbb{I}$ is the identity operator. EUR with quantum memory is called quantum-memory-assisted entropic uncertainty relation (QMA-EUR). QMA-EUR has the wide range of application such as entanglement detection \cite{Huang,Prevedel,Chuan-Feng} and quantum cryptography \cite{Tomamichel,Ng}. Many works have been done to tight the uncertainty bound of QMA-EUR \cite{Pati,Pramanik,Coles,Liu,Zhang,Pramanik1,Adabi,Adabi1,Dolatkhah,Jin-Long}. However, so far the tightest bound is provided by Adabi et al. \cite{Adabi}. According to their results QMA-EUR is given by
\begin{equation}\label{concen}
S(\hat{Q}\vert B)+S(\hat{R}\vert B) \geq \log_2 \frac{1}{c} + S(A \vert B)+ \max \lbrace 0, \delta\rbrace,
\end{equation}
where 
\begin{equation}
\delta = I(A;B)-(I(\hat{Q};B)+I(\hat{R};B)).
\end{equation}
Where $I(\hat{P};B))$ is the Holevo quantity. When Alice measures observable $\hat{P}$ on her particle,  the i-th outcome is obtained with probability $p_i= tr_{AB}(\Pi_i^{A} \rho^{AB} \Pi_i^{A})$ and post measurement state of the Bob is obtained as $\rho_i^{B}=\frac{tr_{A}(\Pi_i^{A} \rho^{AB} \Pi_i^{A})}{p_i}$. Then we have the following relation for Holevo quantity $I(\hat{P};B))$
\begin{equation}\label{holevo}
I(\hat{O};B)=S(\rho^{b})-\sum_{i}p_{i}S(\rho_{i}^{B})
\end{equation}
which is equal to upper bound of Bob's accessible information about Alice's measurement outcomes.

In the real world each quantum system can interacts with its surrounding, so the study of open quantum systems has the special significance\cite{Breuer}. In the most recent work, it has been assumed that  the quantum memory $B$ have an interaction with environment.  In this situation the effects of environmental parameter on the bound of QMA-EUR is studied\cite{Zhang1,Zou,Karpat,Wang,Wang1,Zhang2,Huang2,Huang3,Wang2,Zhang3,Chen,Haseli3,Li,Bai,Hu,chen2,Haseli5}. The bound of QMA-EUR can be control in a dissipative environment by using the quantum weak measurements and measurement reversal \cite{Haseli2,Zhang4}. In this work we will consider the situation that the quantum memory moves inside  a leaky cavity. We will show that the lower bound of QMA-EUR is decreased  by increasing the velocity of the moving quantum memory inside leaky cavity.
\section{Model}
\begin{figure}[!]  
\centerline{\includegraphics[scale=0.7]{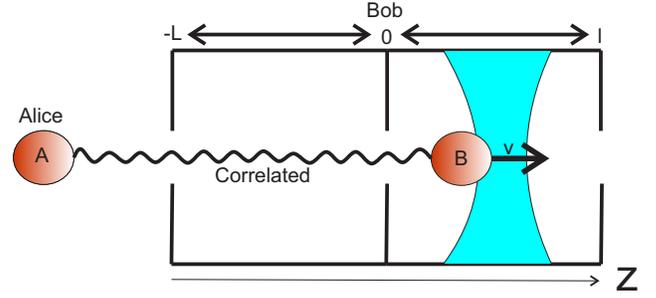}}
\caption{Schematic representation of the model where quantum memory moves inside leaky cavity. While Alice performs measurement on her particle. }\label{Fig1}
\end{figure}

Let us consider the quantum system consist of a two level quantum system and structured environment. The a structure of the environment consists of two mirrors at $z=-L$ and $z=l$ along with partially reflecting mirror in $z=0$. It actually forms a sort of two sequential cavities ($-L,0$) and ($0,l$). The structure of the environment is sketched in Fig.(\ref{Fig1}). 

The classical electromagnetic field $E(z,t)$ in ($-L,l$)  can be written as 
\begin{equation}
E(z,t)=\sum_k E_k(t)U_k(z)+ E_k^{*}(t)U_k^{*}(z),
\end{equation}
where $E_k(t)$ is the amplitude in the $k$-th mode and $U_k(z)$'s are exact monochromatic modes at frequency $\omega_k=ck$\cite{Lang,Gea,Leonardi}. It is assumed that the  electromagnetic field  is polarized along the $x$-direction inside the cavity. The mode functions $U_k(z)$ is given by
\begin{equation}
U_k(z)=
\begin{cases}
  \varepsilon_k \sin k(z+L)  & z < 0;   \\ \\
 J_k \sin k(z-L) & z>0,
 \end{cases} 
\end{equation}
where $\varepsilon$ gives the values $1$ , $-1$ going from each mode to the subsequent one.   $J_k$ for good cavity is  given by
\begin{equation}
J_k=\frac{\sqrt{c \lambda^{2}/l}}{\sqrt{(\omega_k^{2}-\omega_n)^{2}+\lambda^{2}}},
\end{equation}
where $\omega_n=n \pi c /l$ is the frequency of the quasi mode, $\lambda$ is the damping of the cavity and determine the spectral width of the coupling, it also quantifies the photon leakage from cavity mirrors.  Let us consider the case in which the qubit only interacts with the second cavity ($0,l$). The qubit also moves along $z$-axis with constant velocity $v$.

The Hamiltonian of the total system can be written as 
\begin{eqnarray}
\hat{H}&=&\omega_0 \vert 0 \rangle_s \langle 0 \vert + \sum_k \omega_k a_k^{\dag}a_k \nonumber \\
&=&\sum_k f_k(z)\left[ g_k \vert 1 \rangle_s \langle 0 \vert a_k + g_k^{*} \vert 0 \rangle _s\langle 1 \vert a_{k}^{\dag}\right] ,
\end{eqnarray}
where $\omega_0$ is the transition frequency, $\vert 0 \rangle_s$  and $\vert 1\rangle_s$ are the excited and ground state of the qubit system, respectively. $a_{k}^{\dag}$ and $a_k$ are creation and annihilation operators, respectively. $g_k$ quantifies the coupling between the qubit and the cavity. The function $f_k(z)$ represent the form of the qubit motion along $z$-axis which is given by\cite{Leonardi,Mortezapour}
\begin{equation}
f_k(z)=f_k(vt)=\sin(k(z-l))=\sin(\omega_k(\beta t -\tau)),
\end{equation}  
where $\beta=v/c$ and $\tau=l/c$. According to the results which have obtained in Ref. \cite{Mortezapour1}, we will choose the parameter $\beta$ as $\beta=(x)\times 10^{-9}$. This value is equivalent to $v=0.3 (x)$ for the $^{85}Rb$ Rydberg microwave qubit . Now, we assume that the initial state of the total system has the following form
\begin{equation}
\vert \psi(0) \rangle = \left[ c_1 \vert 0 \rangle_s + c_2 \vert 1 \rangle_s \right] \otimes \vert 0 \rangle_c,
\end{equation}
where $\vert c_1 \vert^{2}+\vert c_2 \vert^{2} =1$ and the cavity mode in the vacuum state $\vert 0 \rangle_c$. Then at time $t$ the quantum state of the system can be written as
\begin{equation}
\vert \psi(t) \rangle = c_1 A(t) \vert 0 \rangle_s \vert 0 \rangle_c + c_2 \vert 1 \rangle \vert 0 \rangle_c + \sum_k B_{k}(t) \vert 1 \rangle_s \vert 1_k \rangle_c,
\end{equation}
where $\vert 1_k \rangle_c$ shows the cavity state containing one excitation in the $k$-th mode. By making use of the schr$\ddot{O}$dinger equation one can obtain 
\begin{eqnarray}
i \dot{A}(t)&=&\omega_0 A(t) +  \sum_k g_k J_k f_k(vt)B_k(t), \label{ss1}\\
i \dot{B}_k(t)&=&\omega_k B_k(t) +g^{*}_k J_k f_k(vt)A(t), \label{ss2}
\end{eqnarray}
Solving Eq. \ref{ss2} and inserting its solution to Eq. \ref{ss1} leads to
\begin{eqnarray}\label{ss3}\small \small
\dot{A}(t)+i\omega_0 A(t) &=&  \\
-&& \int_0^{t} dt^{\prime}A(t^{\prime})\sum_k \vert g_k \vert^{2}J_k^{2}f_k(vt)f_k(v t^{\prime})e^{-i\omega_k(t-t^{\prime})}.\nonumber 
\end{eqnarray}
By succession the probability amplitude as $A(t^{\prime})=\tilde{A}(t^{\prime})e^{i \omega_0 t^{\prime}}$, Eq. (\ref{ss3}) can be rewritten as 
\begin{equation}\label{ss11}
\dot{\tilde{A}}+\int_{0}^{t}dt^{\prime}F(t,t^{\prime})\tilde{A}(t^{\prime})=0
\end{equation}
where the memory kernel $F(t,t^{\prime})$ has the following form
\begin{equation}
F(t,t^{\prime})=\sum_k \vert g_k \vert^{2}J_k^{2}f_k(vt)f_k(vt^{\prime})e^{-i(\omega_k-\omega_0)(t-t^{\prime})}.
\end{equation}
The kernel function in the  continuous limit is given bye
\begin{equation}
F(t,t^{\prime})=\int_0^{\infty} J(\omega_k)f_k(t,t^{\prime})e^{-i(\omega_k-\omega_0)(t-t^{\prime})}d\omega_k,
\end{equation}
where $f_k(t,t^{\prime})=\sin(\omega_k(\beta t -\tau))\sin(\omega_k(\beta t^{\prime} -\tau))$ and $J(\omega_k)$ is the spectral density of the electromagnetic field inside the cavity. Here we assume that the spectral density has the Lorentzian form 
\begin{equation}
J(\omega_k)=\frac{1}{2 \pi}\frac{\gamma \lambda^{2}}{(\omega_n-\omega_k)^{2}+\lambda^{2}},
\end{equation}
where $\omega_k$ is the frequency of $k$-th cavity mode and $\omega_n$ is the center frequency of the cavity modes. The spectral width of the coupling $\lambda$ is related to the correlation time of the cavity $\tau_c$ via $\tau_c=1/\lambda$. The parameter $\gamma$ is connected to the time scale $\tau_s$, over
which the state of the system changes, by $\tau_s=1/\gamma$. In the continuous limit ($\tau \longrightarrow \infty$) when $t>t^{\prime}$ we have
\begin{equation}\label{ss15}
F(t,t^{\prime})=\frac{\gamma \lambda}{4}\cosh\left[\theta(t-t^{\prime})  \right]e^{\bar{\lambda}(t-t^{\prime})}, 
\end{equation}
where $\bar{\lambda}=\lambda-i \Delta$ and $\theta = \beta(\bar{\lambda}+i \omega_0)$. Inserting Eq.(\ref{ss15}) to Eq.(\ref{ss11}) and using Bromwich integral formula on can obtain $\tilde{A}(t)$ as 
\begin{eqnarray}\label{ss16}
\tilde{A}(t)&=&\frac{(q_1+u_+)(q_1+u_-)}{(q_1-q_2)(q_1-q_3)}e^{q_1\gamma t} \nonumber \\
&-&\frac{(q_2+u_+)(q_2+u_-)}{(q_1-q_2)(q_2-q_3)}e^{q_2\gamma t} \nonumber \\
&+&\frac{(q_3+u_+)(q_3+u_-)}{(q_1-q_3)(q_2-q_3)}e^{q_3\gamma t},
\end{eqnarray}
where $q_i$'s ($i=1,2,3$) satisfy the following cubic equation as 
\begin{equation}
q^{3}+2(y_1-iy_3)q^2+(u_+u_-+\frac{y_1}{4})q+\frac{y_1(y_1-iy_3)}{4}=0
\end{equation}
where $y_1=\lambda/\gamma$, $y_2=\omega_0/\gamma$, $y_3=(\omega_0-\omega_n) /\gamma$ and $u_\pm=(1\pm\beta)\pm i\beta y_2-i(1\pm \beta)y_3$. So, from Eq.(\ref{ss16}), one can obtain the ecolved density matrix as 
\begin{equation}
 \rho(t)=\left(
\begin{array}{cc}
 \vert c_1 \vert ^{2} \vert A(t) \vert ^{2}  & c_1c^{*}_2 A(t) \\
 c_2c^{*}_1 A^{*}(t) &  1-\vert c_1 \vert ^{2} \vert A(t) \vert ^{2}  \\
\end{array}
\right).
\end{equation}  
\section{Lower bound of QMA-EUR with moving quantum memory inside a leaky cavity }
In this section we provide our model to study the dynamics of the lower bound of QMA-EUR. Bob prepares the correlated two  qubit state $\rho^{AB}$. Then, he sends one part to Alice and keeps other as a  quantum memory. In our model the  quantum memory  $B$  moves inside a leaky cavity. So, the evolution of the  prepared state $\rho^{AB}$ can be characterize  by local dynamical map $\Lambda_t$, such that 
\begin{equation}\label{dynamics}
\rho^{AB_t}=(\mathbb{I}\otimes \Lambda_t)\rho^{AB}.
\end{equation}
Then, Alice and Bob reach an agreement on the measurement of two incompatible observables which is done by Alice on her part.  Alice declares her choice of the measurement to Bob who wants to guess the Alice's measurement outcomes with a better accuracy. Due to the interaction between quantum memory $B$ with leaky cavity the correlation between measured particle $A$ and quantum memory $B$ will decrease. Due to the inverse relation between correlation and the lower bound of QMQ-EUR, the lower bound will increase. In our model we will examine the effect of moving the quantum memory inside a leaky cavity on the lower bound of QMA-EUR.  Schematic representation of our model for QMA-EUR with moving quantum memory inside a leaky cavity is sketched in Fig. (\ref{Fig1}). 
\subsection{Examples}
\subparagraph{Maximally entangled state:} As the first example we consider the case in which Bob prepares a maximally entangled pure state $\vert \phi \rangle^{AB}=1/\sqrt{2}(\vert 0 0 \rangle_{AB} + \vert 11 \rangle_{AB})$. We assumed that the quantum memory $B$ moves inside leaky cavity while Alice particle does not change. The dynamic of the prepared maximally entangled state is given by
\begin{equation}
 \rho^{AB}(t)=\frac{1}{2}\left(
\begin{array}{cccc}
 \vert A(t) \vert^{2}  & 0 & 0 & A(t) \\
 0 &  1-\vert A(t) \vert ^{2} &0&0  \\
 0 & 0 & 0 & 0 \\
 A^{\star}(t) & 0 & 0 & 1 \\
\end{array}
\right).
\end{equation}  
Now, Alice and Bob reach an agreement on measurement of two observables $\hat{\sigma_x}$ and $\hat{\sigma_z}$. The lower bound of QMA-EUR in Eq.(\ref{concen}) is obtained as 
 \begin{equation}
 U_R=1+S_{bin}(\frac{1-\vert A(t) \vert^{2}}{2})-S_{bin}(\frac{\vert A(t) \vert^{2}}{2})+\max\lbrace0,\delta\rbrace,
 \end{equation}
 where $S_{bin}(x)=-x \log_2 x -(1-x)\log_2(1-x)$ and
 \begin{eqnarray}
 \delta&=&-\frac{1}{2}-\frac{1-\eta}{2}\log_2 (\frac{1-\eta}{4})-\frac{1+\eta}{2}\log_2 (\frac{1+\eta}{4}) \nonumber \\ 
 &-&\frac{\vert A(t) \vert^{2}}{2}\log_2 \frac{\vert A(t) \vert^{2}}{2}-\frac{1-\vert A(t) \vert^{2}}{2}\log_2 \frac{1-\vert A(t) \vert^{2}}{2} \nonumber \\
 &-&S_{bin}(\frac{1-\vert A(t) \vert^{2}}{2})-S_{bin}(\frac{\vert A(t) \vert^{2}}{2}),
 \end{eqnarray}
 with $\eta = \sqrt{1-\vert A(t) \vert^{2}+\vert A(t) \vert^{4}}$.
 
 \begin{figure}[!]  
\centerline{\includegraphics[scale=0.4]{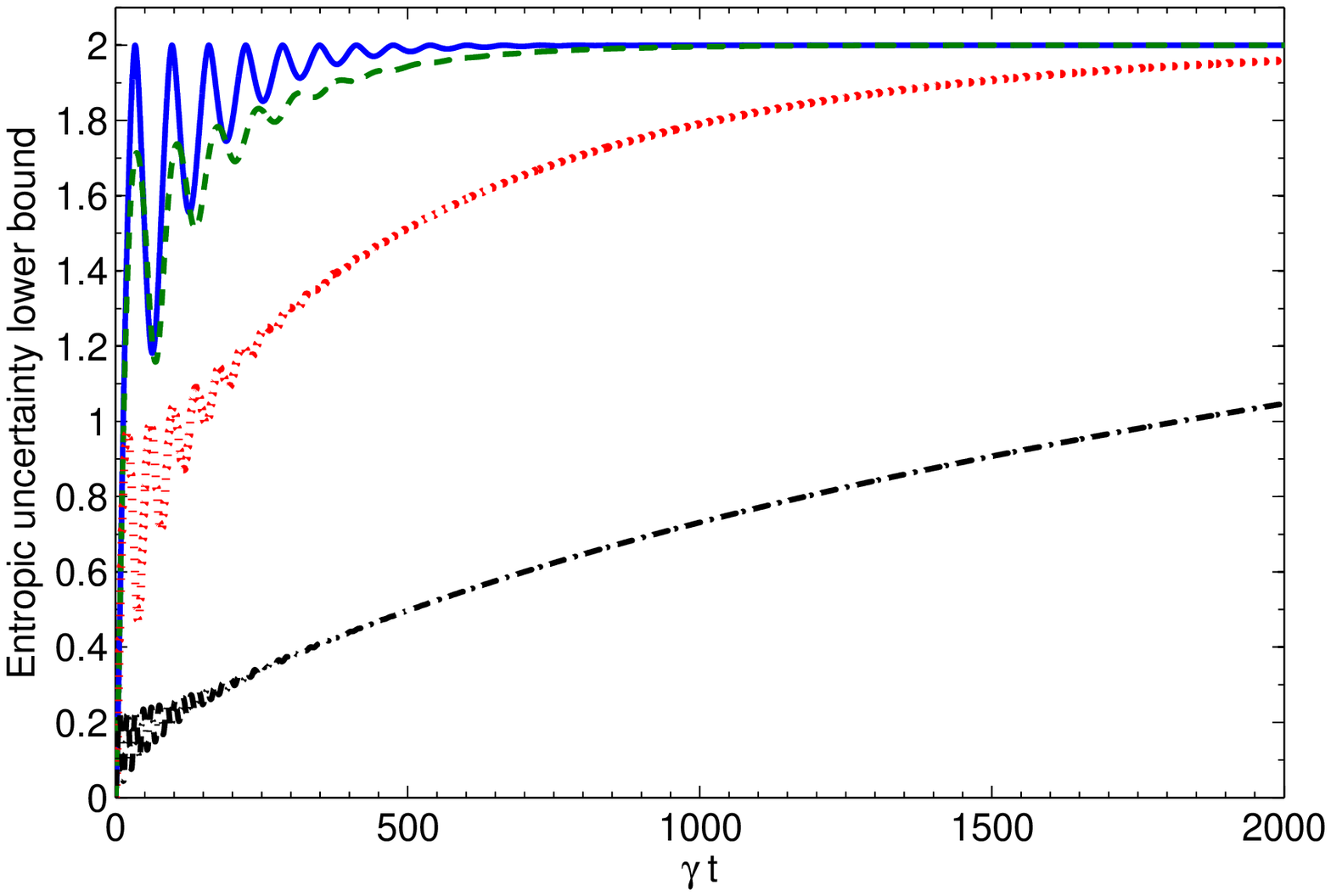}}
\caption{Entropic uncertainty lower bound with moving quantum memory inside a leaky cavity for different value of the speed of quantum memory in non-Markovian regime $\lambda=0.01 \gamma_0$ and $\omega_0=\omega_n=1.53 GHz$, for maximally entangled prepared state. (Solid blue line)$\beta=0$, (dashed green line)$\beta=0.05 \times 10^{-9}$, (dotted red line)$\beta=0.1 \times 10^{-9}$, (dotted dashed black line)$\beta=0.3 \times 10^{-9}$}\label{Fig2}
\end{figure}
\begin{figure}[!]  
\centerline{\includegraphics[scale=0.4]{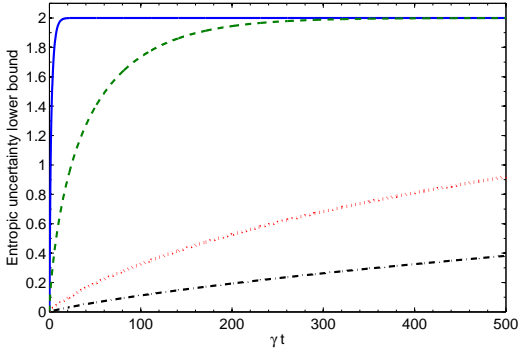}}
\caption{Entropic uncertainty lower bound with moving quantum memory inside a leaky cavity for different value of the speed of quantum memory in Markovian regime $\lambda=3 \gamma_0$ and $\omega_0=\omega_n=1.53 GHz$, for maximally entangled prepared state. (Solid blue line)$\beta=0$, (dashed green line)$\beta=10 \times 10^{-9}$, (dotted red line)$\beta=50 \times 10^{-9}$, (dotted dashed black line)$\beta=100 \times 10^{-9}$}\label{Fig3}
\end{figure}
 In Fig. (\ref{Fig2}), the lower bound of QMA-EUR, $U_R$ is shown as a function of time in non-Markovian regime $\lambda=0.01 \gamma_0$ for different value of the quantum memory speed $\beta$ inside a leaky cavity. From Fig. (\ref{Fig2}), one can see because of the interaction between moving quantum memory and leaky cavity, EULB is increased through time . It is obvious that the lower bound decreases by increasing the speed of the quantum memory $B$. In Fig. (\ref{Fig3}),  $U_R$ is represented as a function of time in Markovian regime $\lambda = 3 \gamma_0$ for different value of the quantum memory speed $\beta$ . From Fig. (\ref{Fig3}), it is obvioused that  $U_R$ is increased over time, while it is decreased by increasing the speed of the quantum memory. 
 
 \subparagraph{Bell diagonal state:} As the second example, we consider the set of two-qubit states
with the maximally mixed marginal states
\begin{equation}
\rho^{AB_i}=p \vert \psi^- \rangle\langle \psi^- \vert + \frac{1-p}{2}(\vert \psi^+ \rangle\langle \psi^+ \vert + \vert \phi^+ \rangle\langle \phi^+ \vert),
\end{equation} 
where  $p \in \left[0,1 \right] $ and  $\vert \phi^{\pm} \rangle = \frac{1}{\sqrt{2}}[\vert 00 \rangle \pm \vert 11 \rangle]$ and $\vert \psi^{\pm} \rangle = \frac{1}{\sqrt{2}}[\vert 01 \rangle \pm \vert 10 \rangle]$ are the Bell diagonal states.   We supposed that the quantum memory $B$ moves inside leaky cavity, while Alice particle does not change. We consider the Bell diagonal state with $p=1/2$ .  The dynamic of the prepared Bell diagonal state is given by
\begin{equation}
\rho_{AB}=\left(
\begin{array}{cccc}
 \rho_{11}^{t} & 0 & 0 & \rho_{14}^{t} \\
 0 & \rho_{22}^{t} & \rho_{23}^{t} & 0 \\
 0 & \rho_{32}^{t} & \rho_{33}^{t} & 0 \\
 \rho_{41}^{t} & 0 & 0 & \rho_{44}^{t} \\
\end{array}
\right)
\end{equation}
with
\begin{eqnarray}
\rho_{11}^{t}&=&\frac{1+p}{4} \vert A(t) \vert^{2}, \quad \rho_{14}^{t}=\frac{1-p}{2} \vert A(t) \vert \nonumber \\
\rho_{22}^{t}&=&\frac{1-p}{4}+\frac{1+p}{4} \left(1- \vert A(t) \vert ^{2} \right), \quad \rho_{23}^{t}=\frac{1-3p}{2}  A(t) \nonumber \\
\rho_{33}^{t}&=&\frac{1}{4} (1-p) \vert A(t) \vert^{2}, \quad \rho_{44}^{t}=\frac{1+p}{4}+\frac{1}{4} (1-p) \left(1-\vert A(t) \vert^{2}\right). \nonumber 
\end{eqnarray}
Now, Alice and Bob reach an agreement on measurement of two observables $\hat{\sigma_x}$ and $\hat{\sigma_z}$. The lower bound of QMA-EUR in Eq.(\ref{concen}) is obtained as
\begin{eqnarray}
U_R&=&1-(a_- -b)\log_2 (a_- -b) - (a_+ -b)\log_2 (z_+ -b) \nonumber \\
&-&(a_- + b)\log_2 (a_- + b)-(a_+ +b)\log_2 (a_+ + b) \nonumber \\
&+& \max \lbrace 0 , \delta \rbrace-S_{bin}(\frac{\vert A(t) \vert ^{2}}{2}),
\end{eqnarray} 
where 
\begin{eqnarray}
a_\pm &=& (2 \pm \vert A(t) \vert ^{2} )/2, \nonumber \\
b&=&(\sqrt{1-\vert A(t)  \vert^{2} + \vert A(t) \vert ^{4}})/4,
\end{eqnarray}
 and 
\begin{eqnarray}
\delta&=&(a_- -b)\log_2 (a_- -b) + (a_+ -b)\log_2 (a_+ -b) \nonumber \\
&+&(a_- + b)\log_2 (a_- + b)+(a_+ +b)\log_2 (a_+ + b) \nonumber \\
&-&S_{bin}(\frac{\vert A(t) \vert ^{2}}{2})+S(\rho_{\hat{\sigma_{z}}B})+S(\rho_{\hat{\sigma_{x}}B}).
\end{eqnarray}
with post measurement von Neumann entropies as
\begin{eqnarray}
S(\rho_{\hat{\sigma_{x}}B})&=&- \frac{\vert A(t) \vert^{2}}{2}\log_2 \frac{\vert A(t) \vert^{2}}{4} \nonumber   \\
&-&\frac{2-\vert A(t) \vert^{2}}{2}\log_2 \frac{2-\vert A(t) \vert^{2}}{4}, \nonumber \\
S(\rho_{\hat{\sigma_{z}}B})&=& - \frac{\vert A(t) \vert ^{2}}{8} \log_2 \frac{\vert A(t) \vert ^{2}}{8} -\frac{3 \vert A(t) \vert ^{2}}{8} \log_2 \frac{3 \vert A(t) \vert ^{2}}{8}\nonumber \\
&-&\frac{4 -3 \vert A(t) \vert ^{2}}{8}\log_2 \frac{4 -3 \vert A(t) \vert ^{2}}{8} \nonumber \\
&-&\frac{4 -\vert A(t) \vert ^{2}}{8}\log_2 \frac{4 -\vert A(t) \vert ^{2}}{8},
\end{eqnarray}
\begin{figure}[!]  
\centerline{\includegraphics[scale=0.4]{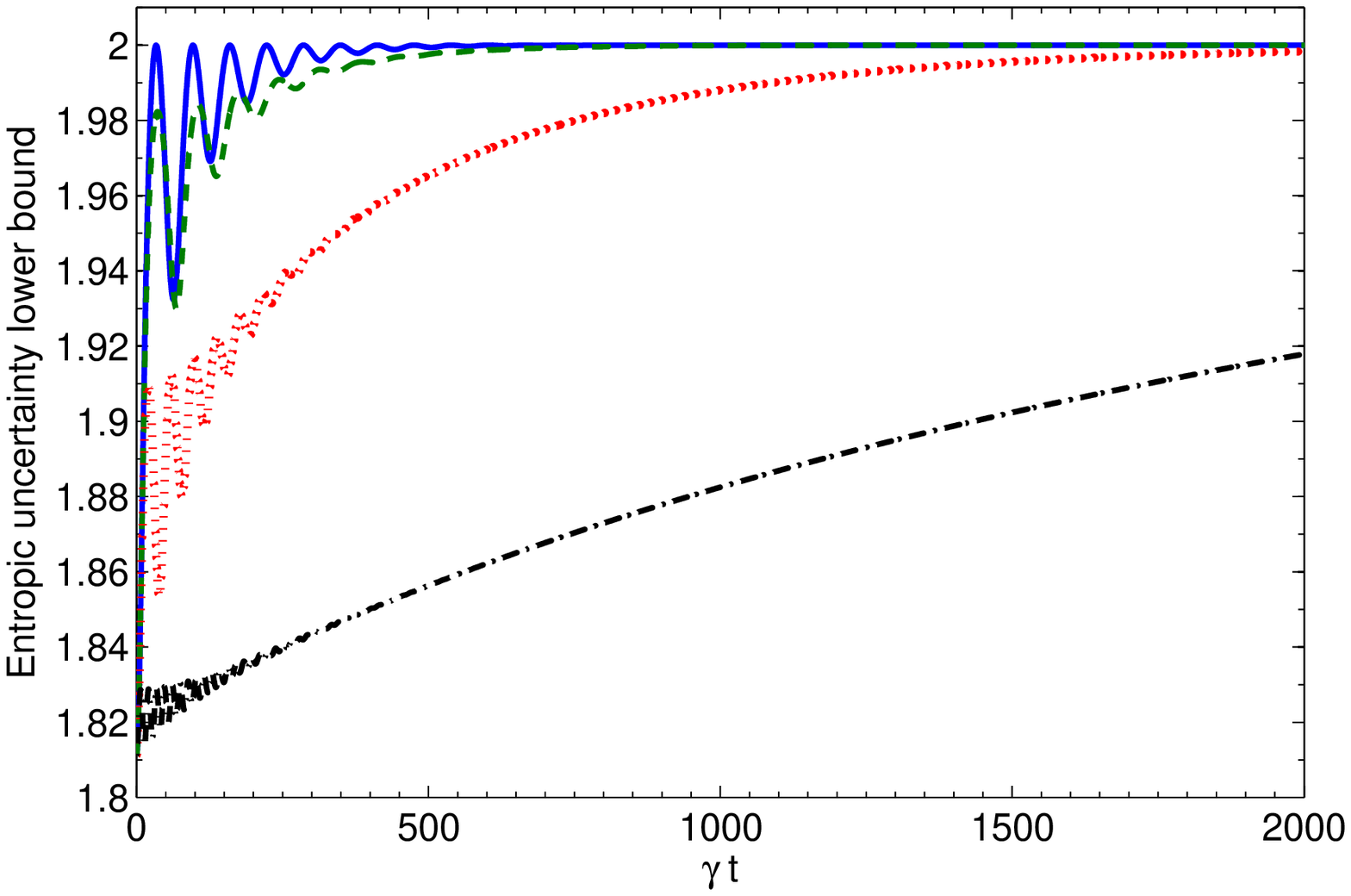}}
\caption{Entropic uncertainty lower bound with moving quantum memory inside a leaky cavity for different value of the speed of quantum memory in non-Markovian regime $\lambda=0.01 \gamma_0$ and $\omega_0=\omega_n=1.53 GHz$, for Bell diagonal prepared state with $p=1/2$. (Solid blue line)$\beta=0$, (dashed green line)$\beta=0.05 \times 10^{-9}$, (dotted red line)$\beta=0.1 \times 10^{-9}$, (dotted dashed black line)$\beta=0.3 \times 10^{-9}$.}\label{Fig4}
\end{figure}
\begin{figure}[!]  
\centerline{\includegraphics[scale=0.4]{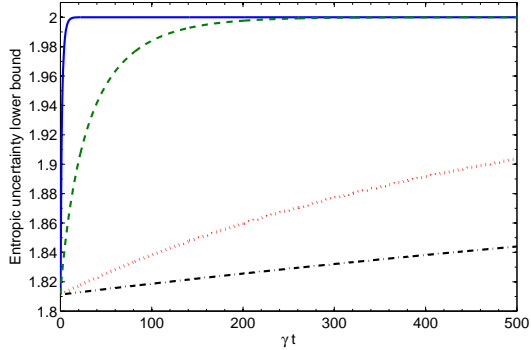}}
\caption{Entropic uncertainty lower bound with moving quantum memory inside a leaky cavity for different value of the speed of quantum memory in Markovian regime $\lambda=3 \gamma_0$ and $\omega_0=\omega_n=1.53 GHz$, for Bell diagonal prepared state with $p=1/2$. (Solid blue line)$\beta=0$, (dashed green line)$\beta=10 \times 10^{-9}$, (dotted red line)$\beta=50 \times 10^{-9}$, (dotted dashed black line)$\beta=100 \times 10^{-9}$. }\label{Fig5}
\end{figure}
 In Fig. (\ref{Fig4}), the lower bound of QMA-EUR, $U_R$ is shown as a function of time for prepared Bell diagonal state  in non-Markovian regime $\lambda=0.01 \gamma_0$ for different value of the quantum memory speed $\beta$ inside a leaky cavity. From Fig. (\ref{Fig4}), one can see because of the interaction between moving quantum memory and leaky cavity, EULB is increased through time . It is obvious that the lower bound decreases by increasing the speed of the quantum memory $B$. In Fig. (\ref{Fig3}),  $U_R$ is represented as a function of time in Markovian regime $\lambda = 3 \gamma_0$ for different value of the quantum memory speed $\beta$ . From Fig. (\ref{Fig4}), it is obvioused that  $U_R$ is increased over time, while it is decreased by increasing the speed of the quantum memory.
 \section{CONCLUSION} 
In this work we have studied the lower bound of QMA-EUR for the case in which the quantum memory part moves inside the leaky cavity. As it is expected the lower bound of QMA-EUR increases due to the interaction between quantum memory and leaky cavity. As it has been shown, the the lower bound of QMA-EUR can be improve by increasing the speed of the quantum memory inside the leaky cavity.

\end{document}